\documentclass[12pt,aps,prb,preprint]{revtex4}
\usepackage{graphicx}
\usepackage{dcolumn}
\usepackage{bm}
\usepackage{amsmath}
\usepackage{textcomp}
\usepackage{mathrsfs}
\usepackage{float}
\usepackage{amssymb}
\usepackage{amsfonts}
\usepackage{mathrsfs}
\usepackage[usenames,dvipsnames]{color}

\begin{document}
%\preprint{}

   \title{Six easy roads to the Planck scale}
    \author{Ronald J. Adler }
     \email{ Electronic mail: gyroron@gmail.com}
     \affiliation{Hansen Laboratory for Experimental Physics Gravity Probe B Mission, Stanford University, Stanford California 94309}

  %\date{\today}

\begin{abstract}
We give six arguments that the Planck scale should be viewed as a fundamental minimum or boundary for the classical concept of spacetime, beyond which quantum effects cannot be neglected and the basic nature of spacetime must be reconsidered. The arguments are elementary, heuristic, and plausible, and as much as possible rely on only general principles of quantum theory and gravity theory. The paper is primarily pedagogical, and its main goal is to give physics students, non-specialists, engineers etc. an awareness and appreciation of the Planck scale and the role it should play in present and future theories of quantum spacetime and quantum gravity.
\end{abstract}

\maketitle

\section{\label{sec:level1}Introduction}

Max Planck first noted in 1899\cite{1,2} the existence of a system of units based on the three fundamental constants,
\begin{align}  %(1)
G&=6.67\times10^{-11} Nm^2/kg^2 (\mbox{or }m^3/kg \mbox{ } s^2)\\
c&=3.00\times10^8m/s\notag\\
h&=6.60\times10^{-34}Js(\mbox{or }kg \mbox{ } m^2/s)\notag
\end{align}
These constants are dimensionally independent in the sense that no combination is dimensionless and a length, a time, and a mass, may be constructed from them. Specifically, using $\hbar \equiv h/2\pi=1.05\times10^{-34} Js$ in preference to $h$, the Planck scale is
\begin{align}  %(2)
l_P&=\sqrt{\frac{\hbar G}{c^3}}=1.6\times10^{-35}m,\mbox{ }T_P=\sqrt{\frac{\hbar G}{c^5}}=0.54\times10^{-43},\\
M_P&=\sqrt{\frac{\hbar c}{G}}=2.2\times10^{-8}kg\notag
\end{align}
The energy associated with the Planck mass is $E_P= M_Pc^2=1.2\times10^{19}GeV$.\cite{3}

The Planck scale is prodigiously far removed from the human scale of about a meter. Indeed we humans are much closer in order of magnitude to the scale of the universe, $10^{26}m$, than to the Planck scale! Present high-energy particle experiments involve energies only of order $10^3GeV$, and even the highest energy cosmic rays detected to date, about $10^{12}GeV$, are far below the Planck energy.

The presence of $\hbar$ in the Planck units in Eq.(2) indicates that the Planck scale is associated with quantum effects, $c$ indicates that it is associated with spacetime, and $G$ indicates that it is associated with gravity. We therefore expect that the scale is characteristic of quantum spacetime or quantum gravity, which is the present conventional wisdom.\cite{3}

We wish to illustrate in this paper that the Planck scale is the boundary of validity of our present standard theories of gravity and quanta. Essentially we travel six roads to the boundary using arguments based on thought experiments. It is beyond our present scope to cross the boundary and discuss current efforts toward theories of quantum gravity and spacetime; however in section IX we will briefly mention a few such efforts,
and we refer the reader to a number of useful references.

Observational confirmation of Planck scale effects is highly problematic.\cite{3,8,9,10} We clearly cannot expect to do accelerator experiments at the Planck energy in the foreseeable future, but there are indirect possibilities. One involves the radiation
predicted by Hawking to be emitted from black holes if they are small enough to have a
``Hawking temperature'' above that of the cosmic background radiation; in the final stages of black hole evaporation the Hawking radiation should have about the Planck energy.\cite{11} Of course Hawking radiation has not yet been observed, although many theorists believe it must exist. Another possibility of interest is to analyze the radiation from very distant gamma ray bursters, which has been enroute for about $10^{10} yr$.\cite{12} Speculations abound on the effect of Planck scale spacetime granularity on propagation of such radiation.\cite{13} Finally we mention that in the earliest stages of cosmological inflation quantum gravity effects might have been large enough to leave an imprint via primordial gravitational radiation on the details of the cosmic microwave background radiation.\cite{14}

Lacking real experiments we use thought experiments (Gedankenexperiment) in this note. We give plausible heuristic arguments why the Planck length should be a sort of fundamental minimum - either a minimum physically meaningful length, or the length at which spacetime displays inescapable quantum properties i.e. the classical spacetime continuum concept loses validity. Specifically the six thought experiments involve: (1) viewing a particle with a microscope; (2) measuring a spatial distance with a light pulse; (3) squeezing a system into a very small volume; (4) observing the energy in a small volume; (5) measuring the energy density of the gravitational field; (6) determining the energy at which gravitational forces become comparable to electromagnetic forces. The analyses require a very minimal knowledge of quantum theory and some basic ideas of general relativity and black holes, which we will discuss in section II. Of course some background in elementary classical physics, including special relativity, is also assumed.

We rely as little as possible on present theory, both because we desire mathematical and pedagogical simplicity, and because the general principles we use are most likely to survive the vagaries of theoretical fashion. We hope that the discussions are thereby made more accessible to physics undergraduates and nonspecialists.

\section{\label{sec:level1}Comments on basic ideas of quantum theory and gravity}

In this section we review the small amount of quantum theory and gravitational theory needed in later sections. Essentially we attempt to compress some root ideas of quantum theory and general relativity into a few paragraphs! Those familiar with quantum theory and general relativity may proceed to the first road in section III.

The first aspect of quantum theory that we recall is the quantization of light into photons. According to Planck and Einstein light of frequency
$\nu$ and wavelength $\lambda = c/\nu$ can only be emitted and absorbed in multiples of the energy
\begin{align}  %(3)
E=h\nu=\hbar\omega
\end{align}
where $\omega$ is the frequency in rad/s. Thus we may think of light as a rain of photons, each with $E = h\nu$.\cite{15,16} According to Einstein¡¦s relation for mass-energy equivalence, $E=mc^2$, a photon should interact gravitationally as if it has an effective mass
\begin{align}  %(4)
M_{ef}=h\nu/c^2
\end{align}
Thus, for example, a single photon captured in a mirrored cavity increases the effective
mass of the cavity according to Eq.(4).

Next we recall a standard heuristic derivation of the Heisenberg uncertainty principle.\cite{15} In Fig.1 we show the Heisenberg microscope thought experiment; using the microscope we view a particle with light entering from the bottom of the page. It is well-known from wave optics (and also rather clear intuitively) that the position of the particle can be determined to an accuracy of about the wavelength $\lambda$ of the light used; a
bit more precisely the uncertainty in position is given by
\begin{align}  %(5)
\Delta x \cong \lambda / sin \vartheta    \label{2.3}
\end{align}
According to classical physics we could determine the position as accurately as desired by using very short wavelength light, of arbitrarily low intensity so as not to disturb the particle by the electric field of the light wave. However the quantization of light as photons with energy $h\nu$ prevents this since the intensity cannot be made arbitrarily low. A single photon scattering from the particle and into the microscope (at angle less than $\vartheta$ ) will impart momentum of order $\Delta p\cong psin\vartheta = (h /\lambda)sin\vartheta$ to the particle, so that Eq.(5) implies

\begin{align}  %(6)
\Delta x \Delta p \cong h \approx \hbar  \label{2.4}
\end{align}
which is the Heisenberg uncertainty principle (UP). (In our rough estimates we do not distinguish between $h$ and $\hbar$, thus taking $2\pi\approx1$; this has been referred to as ``using Feynman units.'') The UP forces us to consider the position and momentum of the
particle to be fundamentally imprecise or ``fuzzy'' in such a way that the particle occupies a region of at least $\hbar$ in phase space
$(x, p)$. Thus we cannot speak of the trajectory of a quantum mechanical particle but must instead take account of the fuzziness of such a
particle in our description, that is in terms of a wave function or probability amplitude.

Most quantum mechanics textbooks also give a derivation of the UP from the commutation relation for the position and momentum operators, and it is also readily obtained from the fact that the wave functions in position and momentum space are Fourier transforms of each other.\cite{16} However neither derivation is as conceptually simple as that using the Heisenberg microscope.

There is an energy-time analog of the UP Eq.(6), but it has a somewhat different meaning. Consider a wave of frequency approximately $\nu$ and duration $T$, which thus consists of $N\approx \nu T$ cycles. The finite duration of the wave means that its leading and trailing edges will in general be distorted somewhat from sinusoidal, so that $N$ will not be precisely well-defined and measurable, but will have an uncertainty of order
$\Delta N \approx1$. This uncertainty in $N$ implies in turn an uncertainty in the frequency, given by $\Delta\nu T\approx \Delta N\approx1$. (This relation is well-known in many fields, such as optics and electrical engineering, wherein it relates band-width and pulse-width;\cite{17} the time
$T$ is subject to a number of somewhat different interpretations. The relation is also easy to derive more formally by calculating the Fourier transform of a finite nearly monochromatic wave train, which is its frequency spectrum.)

Since the energy of a photon of light is given by $E= h\nu=\hbar\omega$ we see that its energy can be measured only to an accuracy given by
\begin{align}  %(7)
\Delta ET\approx h \approx \hbar  \label{2.5}
\end{align}
where $\Delta E$ is, of course, the absolute value of the energy uncertainty. The same relation holds by similar reasoning for most any quantum system.

The expression Eq.(7) formally resembles the UP in Eq.(6), but unlike the position of a particle, time is not an observable in quantum mechanics, so Eq.(7) has a different meaning: $T$ is the characteristic time of the system (eg. pulse width), and not an uncertainty in a time measurement.\cite{18} Thus, for example the light emitted by an atom is only approximately monochromatic, with an energy uncertainty given by $\Delta E\approx\hbar/T$ where $T$ is the lifetime of the excited atomic state.

The quantization relation Eq.(3) and the uncertainty relations Eq.s(6) and (7) are sufficient for our analyses in later sections; we need not discuss detailed quantum mechanics or quantum field theory.

We next move on to general relativistic gravity and black holes. Recall that the
line element or metric of special relativity gives the spacetime distance between nearby events. It is usually expressed in Lorentz coordinates $(ct,\vec{x})$ as\cite{19}
\begin{align}  %(8)
ds^2=(cdt)^2-d\vec{x}^2  \label{2.6}
\end{align}
In general relativity gravity is described by allowing spacetime to be warped or distorted, or, more technically correct, curved. Coordinates in general relativity merely label the points in spacetime and do not by themselves give physical distances; for that we need a metric, which relates coordinate intervals to physical distance intervals.\cite{3,20,21,22} For a weak gravitational field and slowly moving bodies the line element or metric is approximately given by the so-called Newtonian limit
\begin{align}  %(9)
ds^2=(1+2\phi/c^2)(cdt)^2-(1-2\phi/c^2)(d\vec{x})^2 \label{2.7}
\end{align}
Here $\phi$ is the Newtonian potential, with the dimensionless quantity $\phi/c^2$ assumed small, and also assumed to go to zero asymptotically at large distances from the source; for example for a point mass $M$ the potential is $\phi=-GM/r$. What this means is that the proper time, or physical clock time, between 2 events at the same space position and separated only by a time coordinate interval $cdt$ is
\begin{align}  %(10)
ds=cdt\sqrt{1+2\phi/c^2} ~~~~\text{(proper time separation)} \tag{10a} \label{2.8a}
\end{align}
while the physical meter stick distance between 2 events separated only by a space coordinate interval $dx$ is
\begin{align}  %(10)
dx\sqrt{1-2\phi/c^2}    ~~~~\text{(space separation)}  \tag{10b}\label{2.8b}
\end{align}
and similarly for the y and z directions.

The Newtonian limit Eq.(9) is a quite good approximation in many cases; for example in the solar system the Newtonian potential is greatest at the surface of the sun, where $2\phi/c^2\approx10^{-6}$. Thus spacetime in the solar system is extremely close to that of special relativity, or flat. Since we are interested only in order of magnitude estimates, we will make free use of the Newtonian limit Eq.(9) as an approximation.\cite{20,21,22,23}

The exact Schwarzschild metric, which describes a black hole or the exterior of
any spherically symmetric body, is given in the same coordinates as Eq.(9) by\cite{24}
\setcounter{equation}{10}
\begin{align}  %(11)
ds^2=\frac{(1-GM/2c^2r)^2}{(1+GM/2c^2r)^2}(cdt)^2-(1+GM/2c^2r)^4d\vec{x}^2 \label{2.9}
\end{align}
This expression is correct only in matter-free space outside the body. At large distances from the body it is approximately
\begin{align}  %(12)
ds^2\approx (1-&2GM/c^2r)(cdt)^2-(1+2GM/c^2r)d\vec{x}^2,\\
GM/c^2r&<<1,~\notag \label{2.10}
\end{align}
which agrees with the Newtonian limit Eq.(9). If the body is small enough so
that $r=GM/2c^2$ lies outside of it then the coefficient of the time term in Eq.(11) vanishes, which means that the proper time $ds$ is zero, and a clock at that position would appear to a distant observer to stop. This radius defines the surface of the black hole. Light cannot escape from the surface since it undergoes a red shift to zero frequency. A black hole of typical stellar mass is about a km in radius.\cite{3}

A minor caveat concerning Eq.(11) is in order. The coordinate $r$ used in Eq.(11) is called the isotropic radial coordinate and is not the same as that usually used for the Schwarzschild metric, the Schwarschild radial coordinate. We will not use the Schwarzschild radial coordinate here (or even define it) but only note that it is asymptotically equal to the isotropic coordinate $r$ in Eq.(11) but differs from it at the black hole surface by a factor of 4; that is the black hole surface is at radius $2GM/c^2$ in the Schwarzschild coordinate; $2GM/c^2$ is widely known as the Schwarzschild radius.\cite{24}

It is generally believed that if enough mass $M$ is squeezed into a roughly
spherical volume of size about $r\approx GM/c^2$ then it must collapse to form a black hole, regardless of internal pressure or other opposing forces; however if the mass is needle or pancake shaped the question of collapse is not yet clearly settled.\cite{25,26}

In summary the lesson to take away from the above paragraphs is that spacetime
distortion is a measure of the gravitational field. Specifically, since the line element gives the physical distance between nearby spacetime points or events, such distances are given roughly by Eq.(10). This corresponds to a fractional distortion given by
\begin{align}  %(13)
\text{Spacetime fractional distortion}\approx\frac{|\phi|}{c^2} \tag{13a}
\end{align}
Alternatively, from Eq.(11), we may say that for a region that is roughly spherical, of size $l$, and contains a mass $M$ the fractional distortion is of order
\begin{align}  %(13)
\text{Spacetime fractional distortion}\approx\frac{GM}{lc^2} \tag{13b} \end{align}
Even for rather strong gravitational fields this generally holds at least roughly. The above results Eq.(13) will be basic to our arguments in future sections.

\section{\label{sec:level1}The generalized uncertainty principle}
\setcounter{equation}{13}
Our first road to the Planck scale is based on the same thought experiment as the Heisenberg microscope used to obtain the UP in section II and shown in Fig.1, but it includes the effects of gravity to obtain a generalization of the UP.\cite{27,28} According to
the UP in Eq.(4) there is no limit on the precision with which we may measure a particle¡¦s position if we allow a large uncertainty in momentum, as would result from using arbitrarily short wavelength light. But this does not take into account the gravitational effects of even a single photon. As noted in section II a photon has energy $h\nu$ and thus an effective mass $M_{ef}=h\nu/c^2=h/c\lambda$, which will exert a gravitational force on the particle. This will accelerate the particle, making the already fuzzy particle position somewhat fuzzier. Using classical Newtonian mechanics we estimate the acceleration and position change due to gravity as roughly
\begin{align}  %(14)
\Delta a_{\text{g}}\approx GM_{ef}/r_{ef}^2=G(h/\lambda c)/r_{ef}^2,\\
\Delta x_{\text{g}} \approx \Delta a_{\text{g}} t_{ef}^2 \approx G(h/\lambda \notag c)(t_{ef}^2/r_{ef}^2)
 \end{align}
where $r_{ef}$ and $t_{ef}$ denote an effective average distance and time for the interaction. The only characteristic velocity of the system is the photon velocity c, so we naturally take $r_{ef}/t_{ef}\approx c$, and obtain for the gravitational contribution to the uncertainty
\begin{align}  %(15)
\Delta x_{\text{g}}\approx Gh/\lambda c^3 \approx (G\hbar/c^3)/\lambda=l_p^2/\lambda  \label{3.2}
 \end{align}

According to the UP Eq.(4) the position uncertainty neglecting gravity is about $\Delta x\approx \hbar/\Delta p$; we add the gravitational contribution in Eq.(15) to obtain a generalized uncertainty principle or GUP,
\begin{align}  %(16)
\Delta x \approx \Big(\frac{\hbar}{\Delta p}\Big) + l_p^2 \Big( \frac{\Delta p}{\hbar}\Big) \label{3.3}
 \end{align}
This same expression for the GUP has been obtained in numerous ways, ranging in sophistication from our very naive Newtonian approach to several versions of string theory;\cite{27} it would thus appear to be a rather general result of combining quantum theory with gravity and may indeed be correct.\cite{28}

Almost needless to say the GUP should be considered a rough estimate for position uncertainty, with the coefficient of $\Delta p/\hbar$ in the second term only being of order of the square of the Planck length, and the whole expression being valid in order of magnitude as we approach the Planck scale from above. For example, we could just as well have factors of $2\pi$ or 10 or $1/\alpha=137 etc$. multiplying $l_P^2$ in Eq.(16). Moreover we add the uncertainties due to the standard UP and the gravitational interaction linearly; we could equally well have taken the root mean square; this makes little difference in our conclusions, as the reader may verify.

From the GUP in Eq.(16) we see that the position uncertainty of a particle has a minimum at $\hbar/\Delta p= l_P$ and is about
\begin{align}  %(17)
\Delta x_{min} \approx 2l_p \label{3.4}
 \end{align}
as shown in Fig.2. This minimum position uncertainty corresponds to a photon of wavelength about $l_P$ and energy $E_P$.

Since we cannot measure a particle position more accurately than the Planck length, the above result suggests that from an operational perspective the Planck length may represent a minimum physically meaningful distance. As such one may plausibly question whether theories based on arbitrarily short distances, such as string theory, really make sense from an operational point of view; spacetime at the Planck scale and below might not be a useful concept.\cite{3,29}

\section{\label{sec:level1}Light ranging}

The next argument uses a thought experiment that is particularly simple
conceptually, and has the virtue that length is defined via light travel time, which is the actual present practice:\cite{30} the definition of a meter is ``the distance traveled by light in free space in 1/299,792,458 of a second.''

Fig.3 shows the experimental arrangement, which we refer to by the generic name light ranging, in analogy with laser ranging. We send a pulse of light with wavelength $\lambda$ from a position labeled A to one labeled B, where it is reflected back to A, and measure the travel time with a macroscopic clock visible from A. Since light is a wave we cannot ask that the pulse front can be much more accurately determinable than about $\lambda$, so there is an uncertainty of at least about $\Delta l_w\approx\lambda$ in our measurement of the length $l$.

If nature were actually classical we could use light of arbitrarily short wavelength and arbitrarily low energy so as not to disturb the system, and thereby measure the distance to arbitrarily high accuracy. However in the real world we cannot use arbitrarily short $\lambda$ since this would put large amounts of energy and effective mass in the measurement region, even if we use only a single photon.\cite{3} According to our comments in section II the energy of the light will distort the spacetime geometry and thus change the length $l$ by a fractional amount
$|\phi|/c^2$ according to Eq.(13). We may estimate the Newtonian potential due to the photon, which is somewhere in the interval $l$, to be about
\begin{align}  %(18)
\phi \approx \frac{GM_{ef}}{l} \approx \frac{G(E/c^2)}{l}\approx\frac{Gh\nu}{c^2l}\approx\frac{G\hbar}{cl\lambda} \label{4.1}
\end{align}
so the spatial distortion is about
\begin{align}  %(19)
\Delta l_{\text{g}} \approx l(\phi/c^2)\approx(G\hbar/c^3)/\lambda=l_p^2/\lambda \label{4.2}
\end{align}

In Fig.3 the letter B labels a point in space, but we could instead take it to be a small body in free fall, which would move during the measurement (actually only during the return trip of the light pulse), and this would also affect the measurement. We have already estimated just such motion in section III; it is given in Eq.(15) by
\begin{align}  %(20)
\Delta x_{\text{g}}\approx l_p^2/\lambda  \label{4.3}
\end{align}
That is the space distortion in Eq.(19) and the motion in Eq.(20) are comparable, and to our desired accuracy we simply write for either effect $\Delta l_{\mbox{g}}\approx l_P^2/\lambda$.

Since we only know the photon position to be somewhere in $l$ we interpret this as an additional uncertainty due to gravity. We add it to the uncertainty due to the wave nature of the light to obtain
\begin{align}  %(21)
\Delta l \approx \Delta l_w + \Delta l_g \approx \lambda + l_p^2/\lambda \label{4.4}
\end{align}
This expression Eq.(21) for the total uncertainty has a minimum at $\lambda=l_P$, where it is equal to $\Delta l\approx2l_P$, so we conclude that the best we can do in measuring a distance using light ranging is about the Planck length. The energy and gravitational field of the photon prevents us from doing better.

A final note is in order. We have here assumed implicitly that we have access to a perfect classical clock for timing the light pulse. However if the clock is assumed to be a small quantum object then there will be a further contribution to the uncertainty due to the spread of the position wave function of the clock during the travel of the light pulse. Some authors suggest that such a quantum clock should be used in the thought experiment, and arrive at a larger uncertainty estimate for light ranging, one involving the size of the system $l$.\cite{9,31} However other authors point out that such a quantum clock may not be appropriate since it could suffer decoherence and behave classically.\cite{10} Predictions of this nature may be testable with laser interferometers constructed as gravitational wave detectors.\cite{32}

\section{\label{sec:level1}Shrinking a volume}

In this thought experiment we shrink a volume containing a mass $M$ as much as possible, until we are prevented from continuing. We assume the volume is intrinsically three-dimensional, about $l$ in all of its spatial dimensions, as shown in Fig.4. A difficulty occurs due to gravity when the size approaches the Schwarzschild radius,
\begin{align}  %(22)
l\approx GM/c^2 \label{5.1}
\end{align}
The system may then collapse to form a black hole as discussed in section II and cannot be made smaller. There is of course no lower limit to this size if we choose an arbitrarily small mass $M$.

A different difficulty occurs due to quantum effects. From the UP the uncertainty in the momentum of the material in the volume is at least of order $\Delta p\approx \hbar/l$. Since the energy in the volume is given by $E^2 = M^2c^4 + p^2c^2$ the uncertainty in the energy is roughly
\begin{align}  %(23)
\Delta E\approx c\Delta p\approx \hbar c/l  \label{5.2}
\end{align}

If $l$ is made so small that this energy uncertainty increases to about
$2Mc^2$ then pairs of particles can be created and appear in the region around the mass M, as shown in Fig.4.\cite{33} The localization is thereby ruined and the volume cannot shrink further. This limit happens when the energy and size are about
\begin{align}  %(24)
Mc^2\approx\Delta E\approx \hbar c /l,~ l\approx \hbar/Mc \label{5.3}
\end{align}
The quantity $\hbar/Mc$ is known as the Compton radius or Compton wavelength of mass $M$.

Our inability to localize a single particle to better than its Compton radius is wellknown
in particle physics. Indeed one fundamental reason that quantum field theory is used in particle physics is that it can describe the creation and annihilation of particles whereas a single particle wave function
cannot.\cite{34}

We now have two complementary minimum sizes for the volume containing a
mass $M$: the Schwarzschild radius dictated by gravity is proportional to $M$, and the Compton radius dictated by quantum mechanics is inversely proportional $M$. The overall minimum occurs when the two are equal, which happens for

\begin{align}  %(25)
l&\approx \hbar /Mc\approx GM/c^2,~M^2\approx\hbar c/G\equiv M_p^2,\\
l&\approx \hbar/M_pc\approx\sqrt{\hbar G/c^3}=l_p\notag \label{5.4}
\end{align}
Thus the combination of gravity and quantum effects creates insurmountable difficulties if we attempt to shrink a volume to smaller than Planck size.

A minor caveat should be repeated here: we have assumed that the volume is effectively 3 dimensional in that all of its dimensions are roughly comparable; if one of the dimensions is much larger or much smaller than the others the region is effectively one or two dimensional and the question of gravitational collapse is less clear, as noted in section II.\cite{25,26}

\section{Measuring properties of a small volume}

For this rather generic thought experiment we use a quantum probe, such as a light pulse, to measure the size, energy content or other physical properties of a volume of characteristic size $l$, as shown in Fig.5. Such properties may in general fluctuate significantly in the time it takes light to cross the volume, so we would naturally want to do the measurement within that time, $T\approx l/c$. For this we need a probe with frequency greater than $c/l$ and energy greater than about $E\approx\hbar(c/l)$.

But there is a limit to how much probe energy $E$ or effective mass $M_{ef}\approx E/c^2=\hbar/lc$ can be packed into a region of size $l$, as we discussed in section II. According to Eq.(13) the fractional distance uncertainty in the volume containing such an effective mass is about
\begin{align}  %(26)
\frac{\Delta l}{l} \approx\frac{|\phi|}{c^2}\approx\frac{1}{c^2}\Big(\frac{GM_{ef}}{l}\Big)=\frac{G}{c^2l}\Big(\frac{\hbar}{lc}\Big)=\frac{G\hbar}{c^3l^2}=\frac{l_p^2}{l^2}
\end{align}
If $l$ is made so small that this approaches 1 then the geometry becomes greatly distorted and the measurement fails, which happens at $l\approx l_P$.

Another way to see the limit effect is to note that the effective mass $M_{ef}= h/cl$ injected into the region by the probe can induce gravitational collapse to form a black hole (as already noted in section V) when the region size approaches the Schwarzschild radius of about
$GM_{ef}/c^2$; this happens for
\begin{align}  %(27)
l\approx\frac{GM_{ef}}{c^2}\approx\frac{G\hbar}{c^3l},~l\approx\sqrt{G\hbar/c^3}\approx l_p \label{6.2}
\end{align}

We thus conclude that any attempt to measure physical properties in a region of about the Planck size involves so much energy that large fluctuations in the geometry must occur, including the formation of black holes and probably more exotic objects such as wormholes.\cite{35} Such wild variations in geometry were first dubbed spacetime foam by J. A. Wheeler; the phrase has become quite popular to express vividly the supposed chaotic nature of geometry at the Planck scale.\cite{36}

\section{Energy density of gravitational field}

This argument is based on the uncertainty in the energy density of the
gravitational field, and field fluctuations that correspond to the uncertainty. Algebraically it resembles somewhat the argument of section VI, but has a different conceptual basis.

We first obtain an expression for the energy density of the gravitational field in Newtonian theory. Consider assembling a spherical shell of radius $R$ and mass $M$ by moving small masses from infinity to the surface, as shown in Fig.6. From Newtonian theory the energy done moving a small mass $dM$ to the surface is
\begin{align}  %(28)
dE=-GMdM/R  \label{7.1}
\end{align}
and the total energy for the assembly is the integral of this energy over the mass, which is
\begin{align}  %(29)
E=-GM^2/2R  \label{7.2}
\end{align}
This binding energy may be viewed as energy in the gravitational field, and is negative because gravity is attractive. To obtain a general expression for the energy density we
assume it is proportional to the square of the gravitational field $\vec{\mbox{g}}=-\nabla \phi$. That is we set
\begin{align}  %(30)
\rho_{\mbox{g}}=\lambda(\nabla \phi)^2
\end{align}
This is in direct analogy with the energy density of the electric field \textemdash\space except of course for the opposite sign! The proportionality constant $\lambda$ can be determined by integrating $\rho_{\mbox{g}}$ in Eq.(30) over the volume between $R$ and infinity in Fig.6 to get the total field energy. Equating this energy expression with the total binding energy given by Eq.(29) we obtain
\begin{align}  %(31)
\lambda(4\pi G^2M^2/R)=-GM^2/2R,~\lambda=-1/8\pi G \label{7.4}
\end{align}
Thus the energy density of the Newtonian gravitational field is written as
\begin{align}  %(32)
\rho_{\mbox{g}}=-(\nabla \phi)^2/8\pi G \label{7.5}
\end{align}

In general relativity the problem of gravitational field energy is notoriously more subtle and complex.\cite{37} This is due to the nonlinearity of the field equations, which in turn is related to the fact that gravity carries energy and is thus a source of more gravity. In this sense gravity differs fundamentally from the electric field, which does not carry charge and thus is not the source of more electric field. For our present purpose we will content ourselves with the rough estimate given by Eq.(32).

We now consider a space region of size $l$ that is nominally empty and free of
gravity, except for fluctuations allowed by the energy-time uncertainty relation Eq.(7). As in section VI we attempt to measure the
gravitational energy in the region in time $l/c$, with accuracy limited to $\Delta E\approx \hbar c/l$. We thus cannot verify that the
region is truly free of gravity, but only that the gravitational energy in the region is no more than about $\Delta E\approx\hbar c/l$.
From Eq.(32), this limit implies the following limiting relation for the Newtonian potential field
\begin{align}  %(33)
l^3(\nabla \phi)^2/8\pi G\approx\hbar c/l \label{7.6}
\end{align}
As a rough estimate $(\nabla\phi)^2\approx(\Delta\phi/l)^2$ , where $\Delta\phi$ is the uncertainty or fluctuation in the nominally zero Newtonian potential field, so from Eq.(33) we obtain
\begin{align}  %(34)
\Delta \phi\approx\sqrt{\hbar c G}/l  \label{7.7}
\end{align}
This fluctuation corresponds roughly to a fractional spacetime distortion given by Eq.(13),
\begin{align}  %(35)
\Delta l/l\approx\Delta \phi /c^2\approx \sqrt{\hbar G/c^3}/l,~ \Delta l \approx \sqrt{\hbar G/c^3}=l_p \label{7.8}
\end{align}
That is, the allowed nonzero value of the energy density of the gravitational field corresponds to Newtonian potential fluctuations and thus metric and distance fluctuations; the distance fluctuations are, once again, about the Planck length.

\section{Equality of gravity and electric forces}

Our final argument characterizes the Planck scale in terms of the mass or energy at which gravitational effects become comparable to electromagnetic effects and thus cannot be ignored in particle theory. The argument is simple to remember and provides a good mnemonic for quickly deriving the Planck mass.

In most situations we encounter gravity as an extremely weak force; for example the gravitational force between electron and proton in a hydrogen atom is roughly 40 orders of magnitude less than the electric force, and can safely be ignored.\cite{3} However if we instead consider two objects of very large mass $M$ (or rest energy) with the electron charge e, then the gravitational and electric forces become equal when
\begin{align}  %(36)
\frac{GM^2}{r^2}\cong\frac{e^2}{r^2},~M^2\cong\frac{e^2}{G} \label{8.1}
\end{align}
The dimensionless fine structure constant is defined by $\alpha\equiv e^2/\hbar c \cong 1/137$, so we may also express Eq.(36) in terms of the Planck mass as
\begin{align}  %(37)
M^2\cong\alpha\frac{\hbar c}{G}=\alpha M_p^2,~M\cong\sqrt{\alpha}M_p\cong\frac{M_p}{12} \label{8.2}
\end{align}
That is equality occurs within a few orders of magnitude of the Planck mass, at least in terms of Newtonian gravity. We may plausibly infer that such equality also occurs when charged massive particles scatter at near the Planck energy.\cite{38}

Quantum electrodynamics (QED), describing the electromagnetic interactions of electrons and other charged particles, ignores gravitational effects. Clearly this is not reasonable for energies near the Planck scale. Thus virtual processes described by loop integrals are clearly not handled correctly since they involve arbitrarily high energies, and indeed most of them diverge.\cite{3,32,33} We may therefore hope that a more comprehensive
theory that includes gravity might be free of such divergences.

Despite the simplistic nature of this section it does hint at the germ of deep ideas. In the standard model of particle physics the electromagnetic and weak forces are unified into the so-called electroweak force, which has been very successful in predicting experimental results; at low interaction energies the weak and electromagnetic forces differ greatly, but at an energy above a few thousand GeV they become comparable.\cite{38,39} They may be viewed as different aspects of a single force rather than as fundamentally different forces. Similarly the strong force is widely believed to become comparable and similarly unified with the electroweak force in some ¡§grand unified theory¡¨ or GUT at energies of about $10^{16}GeV$, only a few orders of magnitude below the Planck energy.\cite{39} Quite roughly speaking then, all the fundamental forces of nature are believed to become comparable near the Planck scale.

\section{Summary and further comments}

We have tried to show that the Planck scale represents a boundary when we attempt to apply our present ideas of quantum theory, gravity, and spacetime on a small scale. To go beyond that boundary, new ideas are clearly needed.

There is much speculation by theorists on such new ideas. Here we will only mention
three (of many) such efforts very superficially, and refer the reader to references 4 to 7.
The first effort involves perturbative quantum gravity, studied for many years by many
authors, notably by Feynman and Weinberg; in perturbative quantum gravity the flat
space of special relativity is taken to be a close approximation to the correct geometry,
and deviations from it are treated in the same way as more ordinary fields such as
electromagnetism. Just as in quantum electrodynamics Feynman diagrams may be
derived to describe the interactions between particles and the quanta of the gravitational
field, called gravitons. The theory has the serious technical drawback that it does not
renormalize in the same way as quantum electrodynamics, and in fact contains an infinite
number of parameters and graviton interactions. Even more importantly it does not truly
address the quantum nature of spacetime. The second and best-known effort involves
super-string theory, or simply string theory, in which the point particles assumed in
quantum field theories are replaced by one-dimensional strings of about Planck size.
String theory purports to describe all particles and interactions, and has been studied
intensively for decades, and is consistent with gravitational theory since it accommodates
a particle with the properties of the graviton, that is zero mass and spin 2. As yet however
there is no experimental or observational evidence that its basic premise is correct. The
third effort, which we may call affine loop gravity, recasts the mathematics of general
relativity in such a way that the fundamental object is not the metric but a mathematical
object called an affine connection, which is analogous to the gauge potentials describing
other non-gravitational fields, such as the vector potential of the electromagnetic field. In affine loop gravity areas and volumes are indeed quantized, and the theory has other
attractive features.

Various authors\cite{6} visualize spacetime as a boiling quantum foam of strange
geometries such as virtual black holes and wormholes, or as a dense bundle of 6
dimensional Calabi-Yau manifords, or as a subspace of a more fundamental 10 or 11
dimensional space, or as a lower dimensional holographic projection, or as the eigenvalue
space of quantum operators, or as a spin network, or as a woven quantum fabric etc. etc.
But despite intense effort over decades none of the many speculative ideas and theories
has yet reached a high level of success or general acceptance, and we remain free to
consider many possibilities.

Perhaps the oddest possibility is that spacetime at the Planck scale is not truly
observable and may thus be an extraneous and sterile concept, much as the luminous
ether of the nineteenth century proved to be extraneous after the advent of relativity and
spacetime \textemdash\space thus obviating decades of theoretical speculation.\cite{3} At present it is certainly not clear what might replace our present concept of spacetime at the Planck scale.

\section*{ACKNOWLEDGEMENTS}
This work was partially supported by NASA grant 8-39225 to Gravity Probe B. Thanks go to Robert Wagoner, Francis Everitt, and Alex Silbergleit and members of the Gravity Probe B theory group for useful discussions and to Frederick Martin for patient reading and comments on the manuscript.

\begin{figure}[H]
\center
   \includegraphics[width=5cm]{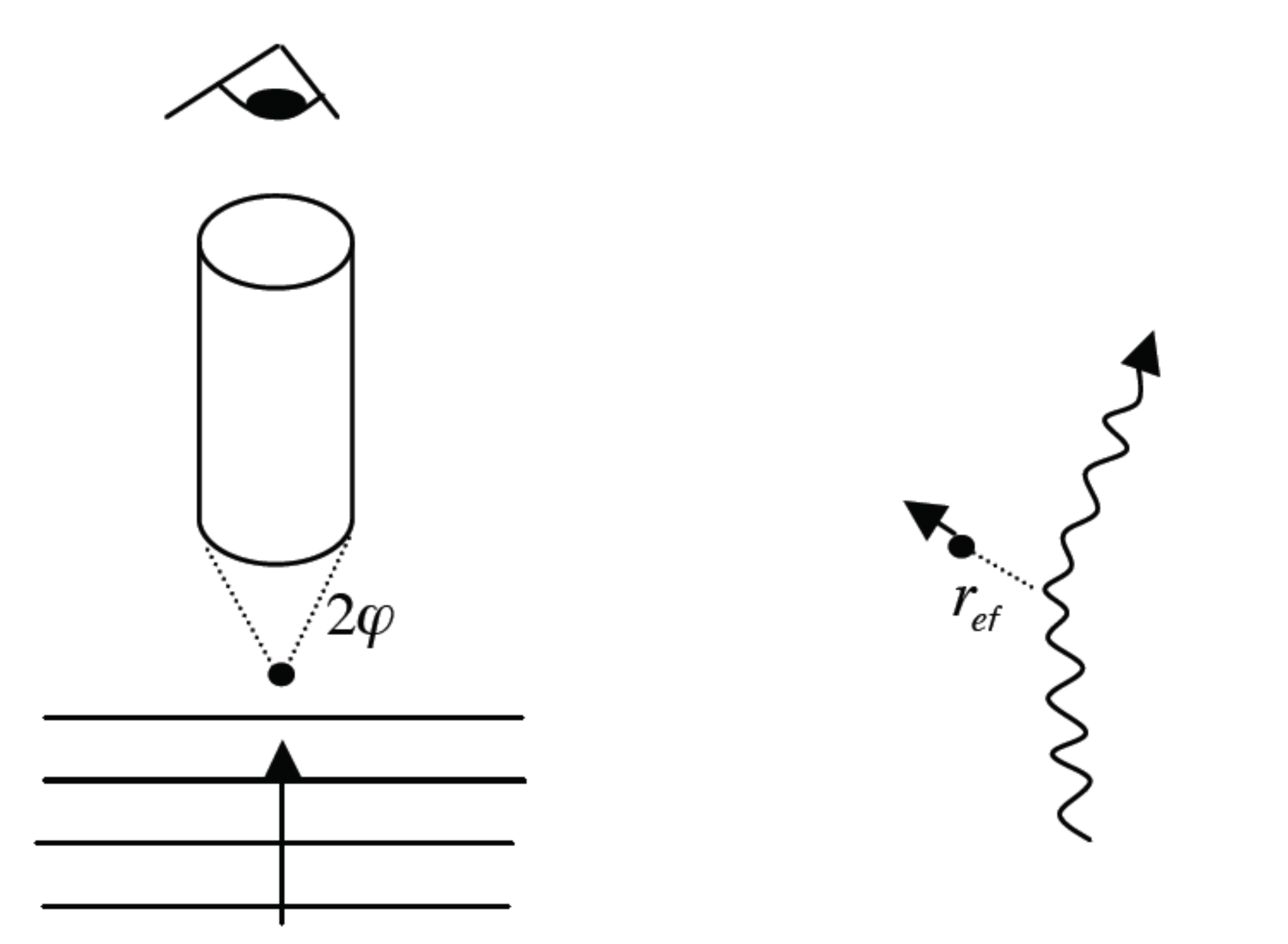}
  \caption{Fig.1. Left: a particle is illuminated from below by light of wavelength $\lambda$, which scatters into the microscope whose objective lens subtends an angle of $2\varphi$. Right: the particle nature of the scattering is emphasized, with an effective separation $r_{ef}$ shown.}
\end{figure}
\begin{figure}[H]
\center
   \includegraphics[width=5cm]{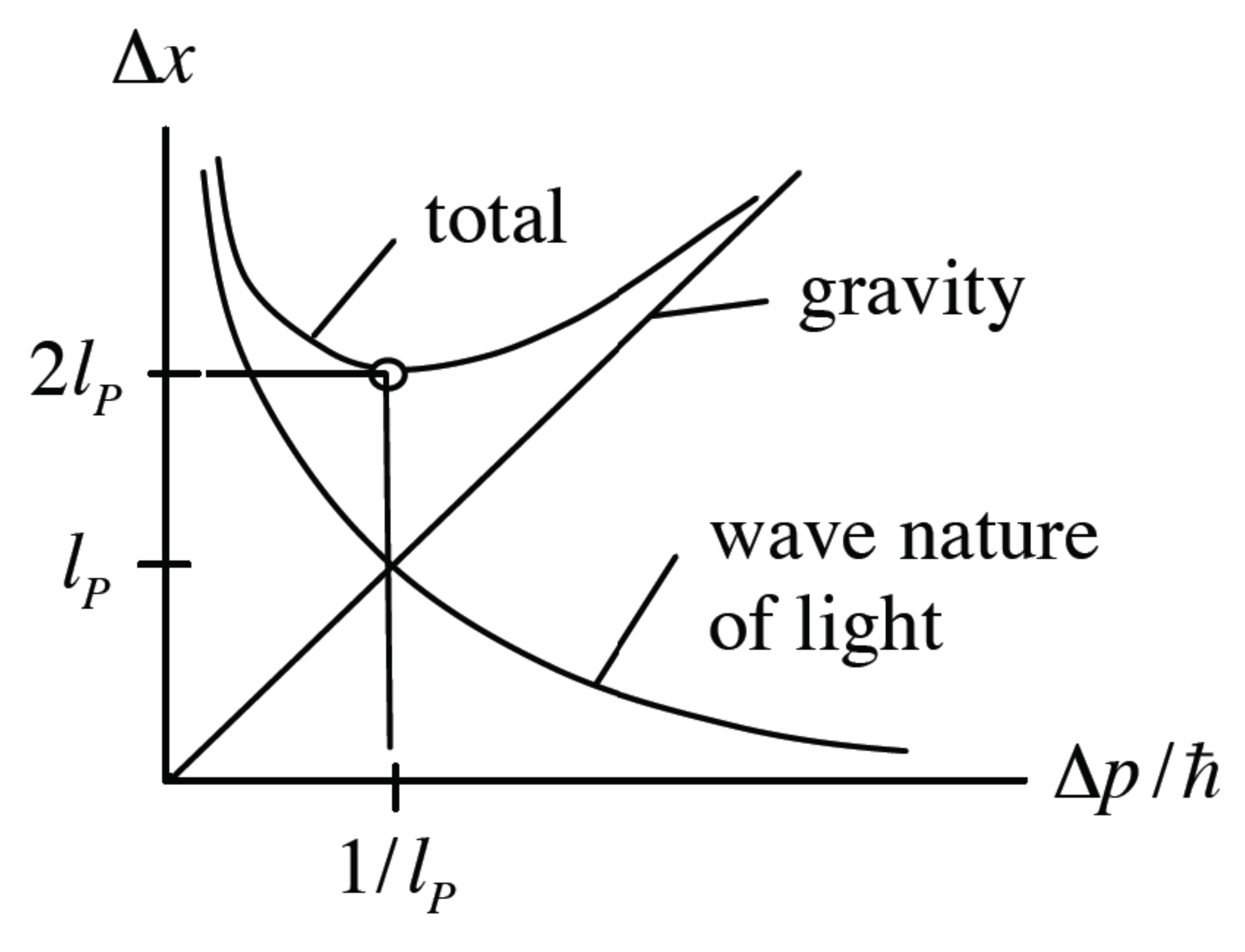}
  \caption{The position uncertainty due to the standard UP and the additional gravitational effect embodied in the GUP. The minimum uncertainty occurs at about twice the Planck length.}
\end{figure}
\begin{figure}[H]
\center
   \includegraphics[width=5cm]{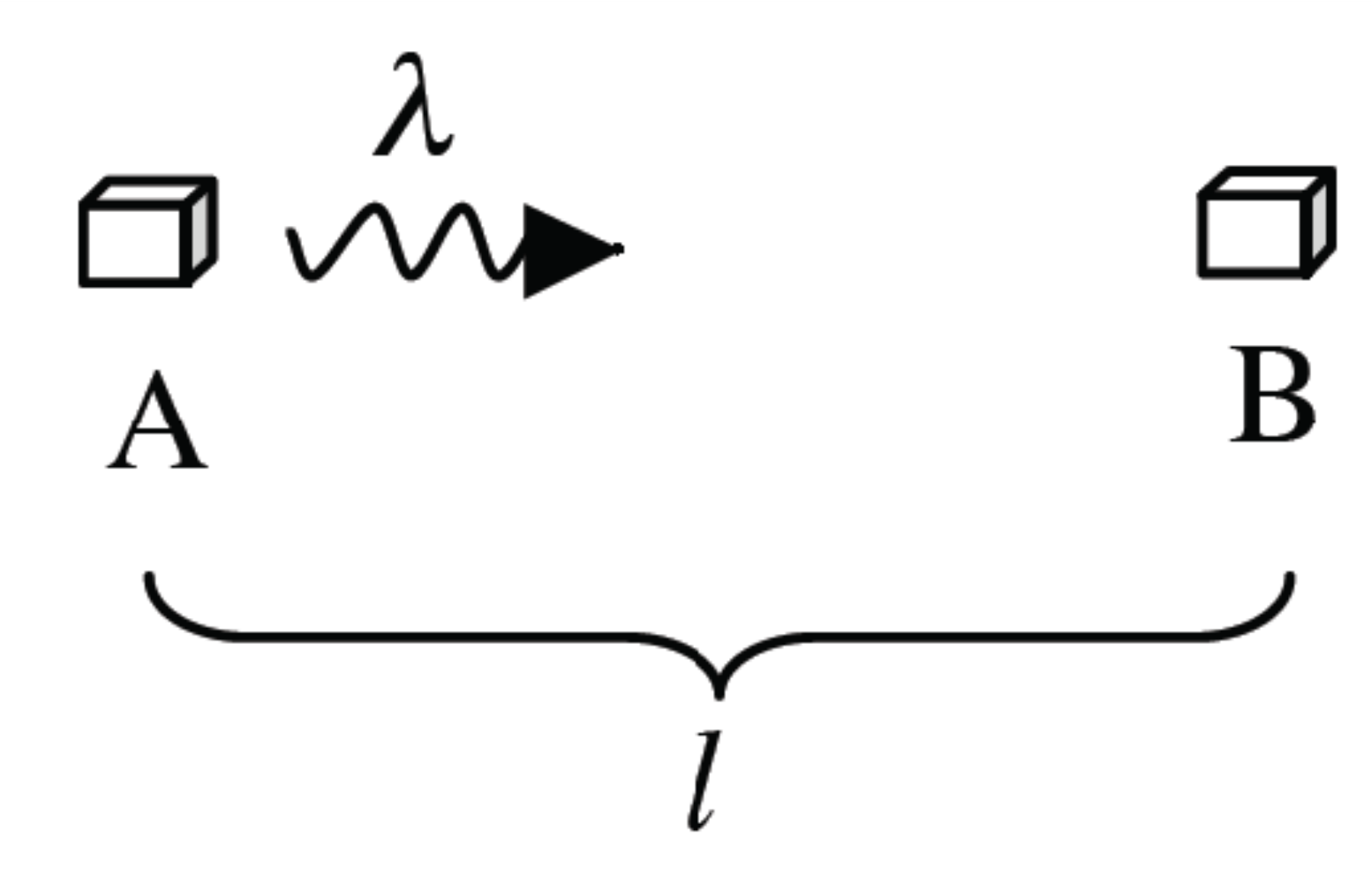}
  \caption{A light pulse is sent from A and reflected back from B. Its energy causes a distortion of the spacetime between A and B and hence affects the length $l$.}
\end{figure}
\begin{figure}[H]
\center
   \includegraphics[width=5cm]{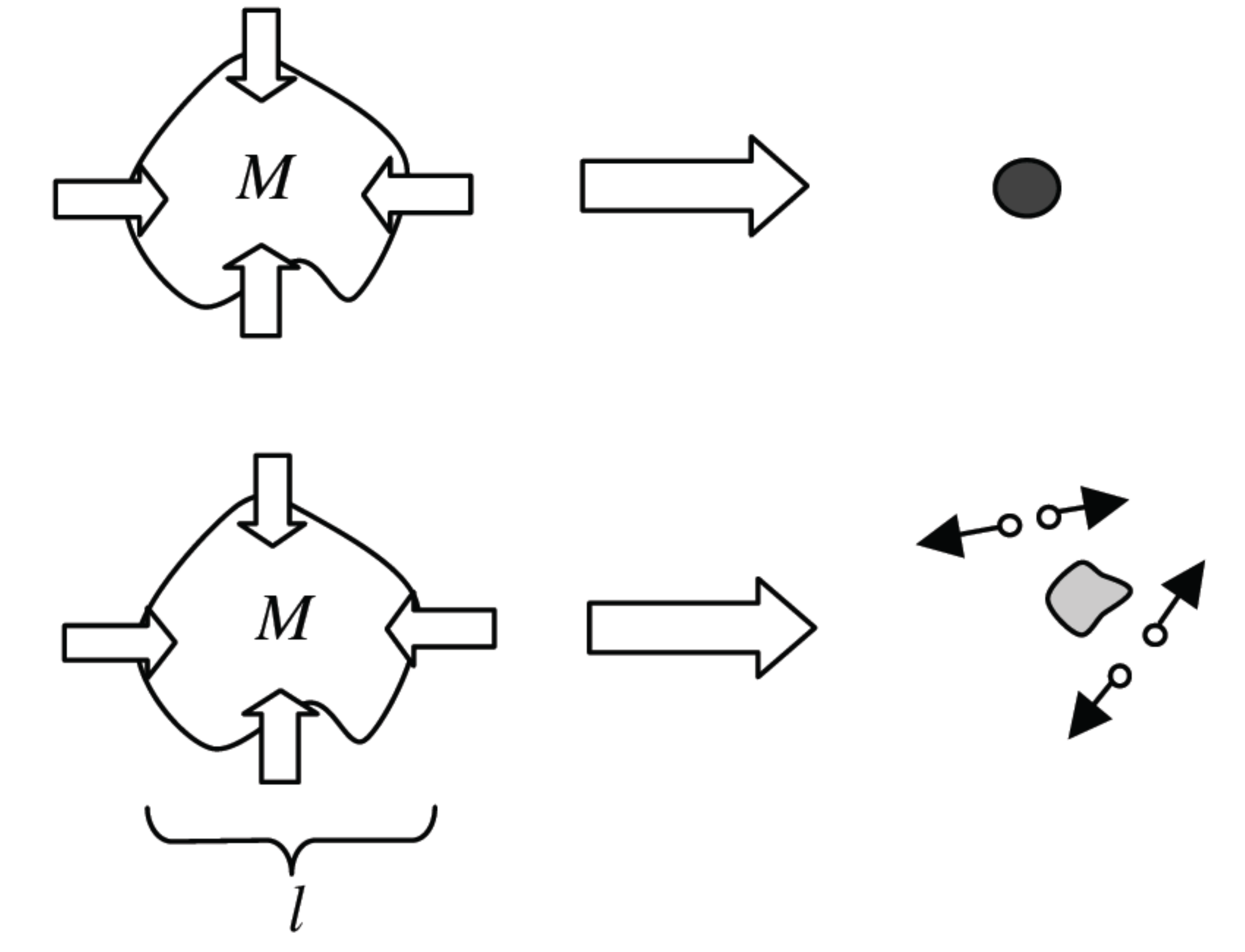}
  \caption{Shrinking of a volume containing mass $M$ is limited by gravitational collapse to a black hole in the top figure, and by the creation of particle anti-particle pairs in the lower figure.}
  \end{figure}
  \begin{figure}[H]
\center
   \includegraphics[width=5cm]{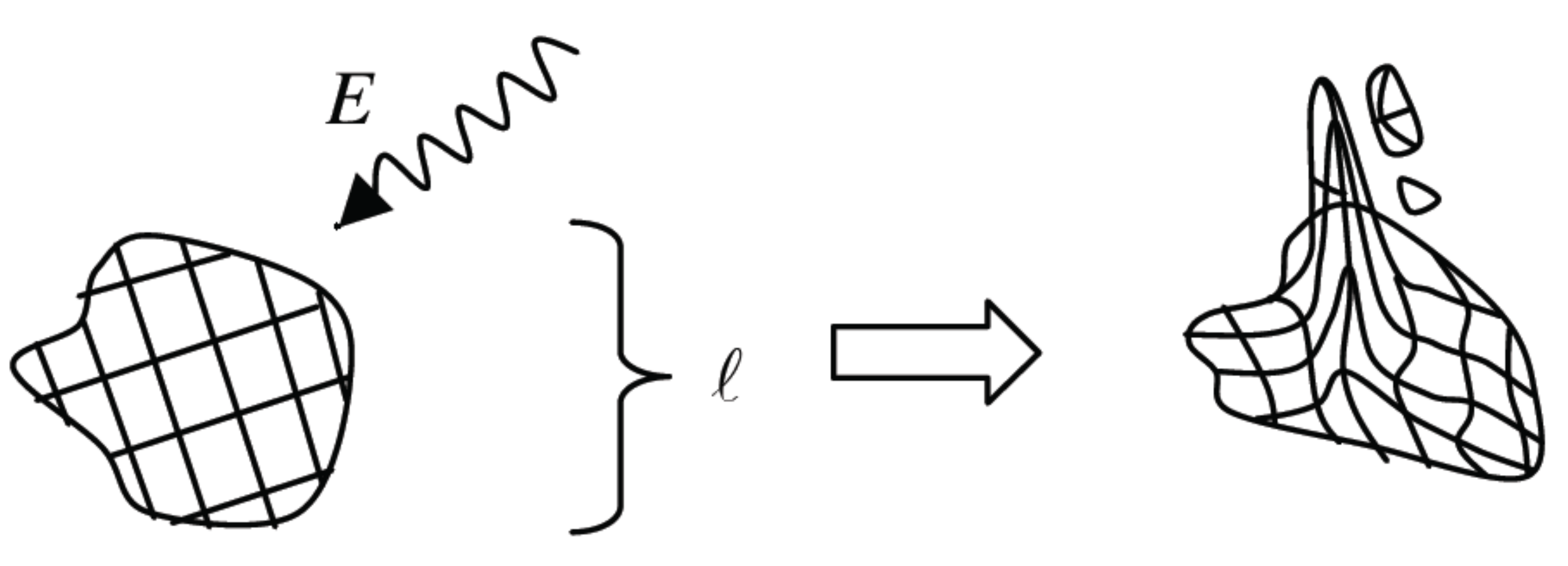}
  \caption{A region of space of size $l$ to be measured in time $l/c$. As the size approaches the Planck length there can occur wild variations in the geometry, including such things as black holes and wormholes.}
\center
   \includegraphics[width=5cm]{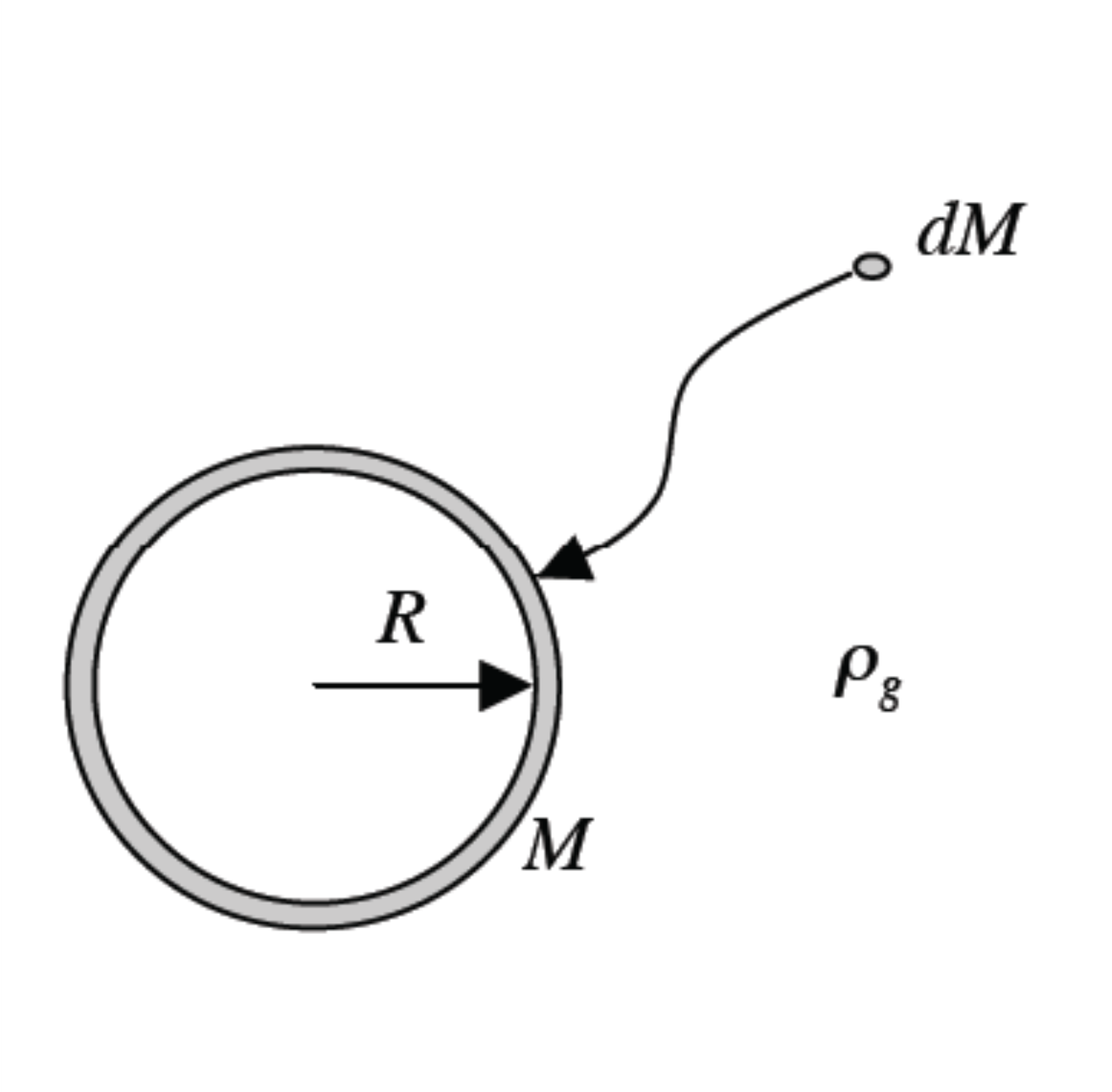}
  \caption{Potential energy of many small $dM$ elements reappears as energy density $\rho_\text{g}$ of the gravitational field.}
\end{figure}
\end{document}